\documentclass[twocolumn,prb,aps,showpacs,floats,floatfix]{revtex4}
\usepackage[final]{graphicx}
\usepackage{calc}

\begin{document}
\title{Quantum disordering of the $111$ state and the compressible--incompressible transition in quantum Hall bilayer systems}
\author{Zlatko Papi\'{c} and Milica V. Milovanovi\'{c}}
\affiliation{Institute of Physics, P.O.Box 68, 11080 Belgrade,
Serbia}
\date{\today}

\begin{abstract}
We systematically discuss properties of quantum disordered states of
the quantum Hall bilayer at $\nu_{T} = 1$. For one of them,
so-called vortex metal state, we find ODLRO (off-diagonal long-range
order) of \emph{algebraic} kind, and derive its transport properties. It is
shown that this state is relevant for the explanation of the
``imperfect" superfluid behavior, and persistent intercorrelations,
for large distances between layers, that were found in experiments.
\end{abstract}

\maketitle \vskip2pc

\section{Introduction}\label{introduction}

Electrons in quantum Hall bilayer systems at total filling factor
$\nu_{T} = 1$ naturally correlate in two different ways due to Pauli
principle and Coulomb interaction. If the layers are sufficiently
far apart, dominant correlations would be those of intralayer kind
because electrons in one layer are unable to sense what is taking
place in the opposite layer. This does not hold, however, in the
limit of small layer separation. Instead, with decreasing $d/l_B$,
the ratio of the distance between layers to the magnetic length, the
correlations between electrons in different layers gain strength and
begin to compete with intralayer correlations. It is the interplay
of those two kinds of correlations that we focus on in this paper.

For the case of prevalent interlayer correlations, there are already
a few theoretical models at hand which provide a satisfactory
description: $111$-state given by Halperin's~\cite{halperin} $111$
wavefunction
$\Psi_{111}=\prod_{i<j}(z_{i\uparrow}-z_{j\uparrow})\prod_{k<l}(z_{k\downarrow}-z_{l\downarrow})\prod_{m,n}(z_{m\uparrow}-z_{n\downarrow}),$
quantum Hall ferromagnet,~\cite{kmoon} condensate of
excitons~\cite{macdonald} or composite
bosons.~\cite{stanic} Nevertheless, both theoretically and
experimentally, it is evident that with increasing $d/l_B$ a quantum
disordering of this state is bound to take place. For example, the
tunneling peak observed by Spielman \emph{et al.}~\cite{spielman} is
indeed sharp and pronounced, but its nature is more that of a
resonance than of the speculated Josephson effect, while the
temperature dependences of Hall and longitudinal resistance in
experiments of Kellogg \emph{et al.}~\cite{kellogg} and Tutuc
\emph{et al.}~\cite{tutuc} do not provide support to the predicted
Berezinskii-Kosterlitz-Thouless (BKT) scenario of a bilayer finite
temperature phase transition.~\cite{kmoon} Deeper understanding of
the regime $d\sim l_B$ is therefore an important, open problem in
the physics of quantum Hall bilayers and strongly correlated
electron systems in general.

Hereinafter we present some results which pertain to quantum
disordering that is believed to take place in the quantum Hall
bilayer at $\nu_{T}=1$. The ground state at $d=0$ is a Bose
condensate well-described by $111$-wavefunction due to Halperin,
while the low-lying excitations are composite bosons i.e. electrons
dressed with one quantum of magnetic flux.~\cite{stanic} The idea of
disordering that we employ is to allow the formation of composite
fermions (i.e. electrons dressed with \emph{two} quanta of magnetic
flux) that coexist with composite bosons.~\cite{srm03} There are two
ways to introduce composite fermions into the Bose condensate and
this will be explained in Sec.~\ref{sect:basic chern-simons}. We
then pursue a phenomenological Chern-Simons transport theory \`{a}
la Drude in order to examine the elementary predictions of those two
model states. In Sec.~\ref{cs teorija za dvosloj} we arrive at an
effective gauge theory for both cases. This enables us to calculate
the correlation functions, modes of low-lying excitations and
characteristic off-diagonal long-range order (ODLRO). We will be
primarily interested in the pseudospin channel of these states. In
one of those, the so-called vortex metal state that we believe may
appear in the bilayer at larger $d/l_{B}$ as a manifestation of
increasing intracorrelations, we derive an \emph{algebraic} ODLRO. In
Sec.~\ref{evolution} we focus on the incompressible region and the
crossover around the critical layer separation. We will argue that
our field-theoretical, homogeneous picture in fact suggests that
vortex metal, if relevant for the strongly-coupled, incompressible
region, may appear only localized, in form of islands in the
background of the superfluid state for smaller $d/l_{B}$.
In Sec.~\ref{further comparison} we give a more thorough
analysis of the experiments on bilayer, addressing especially
the compressible, weakly-coupled region, and the question of persistent
intercorrelations \cite{kelloggdrag} in the framework of the vortex
metal state. Sec.~\ref{conclusion} is devoted to discussion and conclusion.
For the sake of clarity and in order to make the text
self-contained, some of the known
results~\cite{srm03,milica-preprint} will be re-derived in this
paper.

\section{Trial wavefunctions for the bilayer}\label{sect:basic chern-simons}

\begin{figure}[htb]
\begin{center}\includegraphics[width=\linewidth]{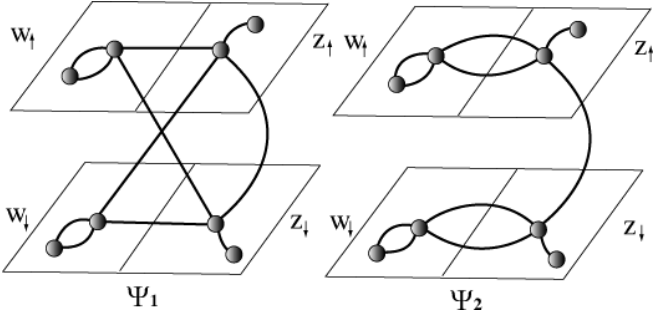} \end{center}
\caption{Correlations between electrons in two layers.}
\label{fig:korelacije}
\end{figure}

Building on Laughlin's proposal for the wavefunction of a single
quantum Hall layer,~\cite{laughlin-orig} the construction of Rezayi-Read
wavefunction~\cite{rezayi-read} for $\nu=1/2$ and Halperin's
$111$-wavefunction for bilayer,~\cite{halperin} we may formally
imagine that there are two species of electrons in each layer
($z,w$) which are all mutually correlated through intracorrelations
(within the same layer) and intercorrelations (between opposite
layers), Fig.~\ref{fig:korelacije}.

Starting from the $111$-function of the Bose condensate, we will
minimally deform it in order to include the composite fermions.
Given that each particle binds the same number of flux quanta and
taking Pauli principle into account, this becomes a combinatorial
problem with two solutions. In the first case:
\begin{eqnarray}\label{psi1}
\nonumber \Psi_1=\mathcal{P}\mathcal{A}\{\prod_{i<j}
(z_{i\uparrow}-z_{j\uparrow})
 \prod_{k<l} (z_{k\downarrow}-z_{l\downarrow}) \prod_{p,q}
 (z_{p\uparrow}-z_{q\downarrow})\\
\nonumber \times \Phi_f(w_\uparrow ,\overline{w}_\uparrow)
\prod_{i<j} (w_{i\uparrow}-w_{j\uparrow})^2 \\
\nonumber \times \Phi_f(w_\downarrow ,\overline{w}_\downarrow)
\prod_{k<l}
(w_{k\downarrow}-w_{l\downarrow})^2\\
\nonumber \times \prod_{i,j}(z_{i\uparrow}-w_{j\uparrow})
\prod_{k,l}(z_{k\uparrow}-w_{l\downarrow}) \\
\times \prod_{p,q}(z_{p,\downarrow}-w_{q,\uparrow})
\prod_{m,n}(z_{m\downarrow}-w_{n\downarrow})
 \}.
\end{eqnarray}
The first line in this formula can be recognized as $111$-function,
followed by two $\nu=1/2$ separate layers ($\Phi_{f}$'s denote the Slater
determinants of free composite fermions), while the last two lines
stem from the flux-particle constraint (all these correlations are
depicted on the left hand side of Fig.~\ref{fig:korelacije}).
$\mathcal{P},\mathcal{A}$ denote projection to the lowest Landau
level (LLL) and fermionic antisymmetrization (independently for each
layer), respectively. In thermodynamic limit, the relation between
the number of particles and flux quanta
reads~\cite{milica-preprint}:
\begin{eqnarray}\label{flux-particle 1}
\nonumber N_{\Phi}=N_{b\uparrow}+N_{b\downarrow}+N_{f\uparrow}+N_{f\downarrow}\\
=2N_{f\uparrow}+N_{b\uparrow}+N_{b\downarrow}=2N_{f\downarrow}+N_{b\uparrow}+N_{b\downarrow}.
\end{eqnarray}
$N_{\Phi}$ is the number of flux quanta through the system, and $N_{b\sigma}$ and $N_{f\sigma}$ 
are the number of bosons and fermions inside the layer $\sigma$ respectively; $\sigma=\uparrow, \downarrow$ is the layer
index. Eq.~(\ref{flux-particle 1}) enforces an additional constraint
$N_{f\uparrow}=N_{f\downarrow}.$ Therefore, the number of fermions
is balanced in two layers, while boson numbers are not subject to
any such constraint. This fact is important
because of the broken symmetry of spontaneous interlayer phase
coherence in the $111$-state, which demands nonconservation of
$N_{b\uparrow}-N_{b\downarrow}$. Although we work in a fixed (relative) number representation (allowed in a broken
symmetry case), to account for a broken symmetry situation we need to have a possibility of unconstrained 
relative number of bosons. Then a superposition of the wavefunctions of the form in Eq.~(\ref{psi1}) would lead to the
usual representation. 

In the second case which is expected to describe dominant
intracorrelations, fermions bind exclusively within the layer they
belong to (right side of Fig.~\ref{fig:korelacije}) and the
corresponding wavefunction is:
\begin{eqnarray}\label{psi2}
\nonumber \Psi_2=\mathcal{P}\mathcal{A}\{\prod_{i<j}
(z_{i\uparrow}-z_{j\uparrow})
 \prod_{k<l} (z_{k\downarrow}-z_{l\downarrow}) \prod_{p,q}
 (z_{p\uparrow}-z_{q\downarrow})\\
\nonumber \times \Phi_f(w_\uparrow ,\overline{w}_\uparrow)
\prod_{i<j} (w_{i\uparrow}-w_{j\uparrow})^2 \\
\nonumber \times \Phi_f(w_\downarrow ,\overline{w}_\downarrow)
\prod_{k<l}
(w_{k\downarrow}-w_{l\downarrow})^2\\
\times \prod_{i,j}(z_{i\uparrow}-w_{j\uparrow})^2
\prod_{k,l}(z_{k\downarrow}-w_{l\downarrow})^2
 \}.
\end{eqnarray}
In this case, the flux-particle relation~\cite{milica-preprint} is:
\begin{eqnarray}\label{flux-particle 2}
\nonumber N_{\Phi}=2N_{f\uparrow}+2N_{b\uparrow}=2N_{f\downarrow}+2N_{b\downarrow}\\
=2N_{f\uparrow}+N_{b\uparrow}+N_{b\downarrow}=2N_{f\downarrow}+N_{b\uparrow}+N_{b\downarrow},
\end{eqnarray}
implying that both fermion and boson numbers must be balanced:
$N_{f\uparrow}=N_{f\downarrow},N_{b\uparrow}=N_{b\downarrow}.$

In Ref.~\onlinecite{srm03} the authors numerically calculated the
overlap of $\Psi_1$ with the exact ground state wavefunction for a
system of $5$ electrons in each layer with varying $d/l_B$. Their
results seem to demonstrate convincingly that (at least for small
systems) the approach with trial wavefunctions that interpolate
between two well-established limits, namely those of $111$-state and
decoupled $\nu=1/2$ layers, is not only an artificial mathematical
construction but corresponds to physical reality. Despite the fact
that the number of electrons in this simulation is certainly well
below the thermodynamic limit, the fact that the overlaps between
$\Psi_{1}$ and the exact ground state display peaks very close to
$1$ at small $d/l_{B}$ provides confidence in the choice of
wavefunction $\Psi_1$ (at least for small $d/l_{B}$).

If there is a phase separation in between the sea of composite
bosons and composite fermions, the phase transition will be of the
first order. Such a scenario is launched in
Ref.~\onlinecite{stern-halperin}, where the authors imagine static,
isolated regions of incoherent phase inside $111$ phase. Although
this model correctly explains some of the observed phenomena (e.g.
semicircle law), the persistence of intercorrelations in the
weakly-coupled, compressible regime \cite{kelloggdrag} which
gradually die out suggests a continuous transition. Such a
possibility is naturally present in the picture of composite
boson-composite fermion mixture.

A transport theory of Drude kind can be easily formulated~\cite{srm03} if we
consider that composite fermions bind two quanta of magnetic flux,
unlike composite bosons which bind only one quantum of magnetic
flux. As long as we remain in the RPA approximation, they can all be
treated as free particles moving in the presence of the effective
field which is given by the sum of the external and
self-consistently induced electric field. In the first case
($\Psi_1$), the effective field as seen by particles in the layer
$\sigma$ is:
\begin{eqnarray} \label{eff_field_1.1}
\mathcal{E}_{f}^{\sigma}=
\mathbf{E}^{\sigma}-2\mathbf{\epsilon}\mathbf{J}_{f}^{\sigma}
-\mathbf{\epsilon}(\mathbf{J}_b^{1}+\mathbf{J}_b^{2}),\\
\label{eff_field_1.2}
 \mathcal{E}_{b}^{\sigma}=
\mathbf{E}^{\sigma}-\mathbf{\epsilon}(\mathbf{J}_{b}^{1}
+\mathbf{J}_b^{2}+\mathbf{J}_f^{1}+\mathbf{J}_{f}^2),
\end{eqnarray}
where $\mathbf{J}_{f(b)}^{\sigma}$ denote Fermi- and Bose-currents
in the layer $\sigma$ and $\mathbf{\epsilon}=\left[ \begin{array}{rrr} 0 & \delta \\
-\delta & 0 \\ \end{array} \right] $, $\delta=\frac{h}{e^2}$.
Transport equations are:
\begin{eqnarray}
\mathcal{E}_{f(b)}^{\sigma}=\rho_{f(b)}^{\sigma}\mathbf{J}_{f(b)}^{\sigma}
\end{eqnarray}
and, as required by symmetry, $\rho_{f(b)}^1=\rho_{f(b)}^2$, while
the total current is given by
$\mathbf{J}^{\sigma}=\mathbf{J}_{b}^{\sigma}+\mathbf{J}_{f}^{\sigma}.$
We define single layer resistance ($\rho^{11}$) and drag resistance
($\rho^D$) as follows::
\begin{eqnarray}\label{definicija otpornosti}
\mathbf{E}^1=\rho^{11}\mathbf{J}^1,\\
\label{definicija draga}
\mathbf{E}^2=\rho^D\mathbf{J}^1.
\end{eqnarray}
When both layers have the same filling, $\nu_1=\nu_2=1/2$, tensors
$\rho_b,\rho_f$ are diagonal (because the composite particles are in
zero net field): $\rho_{b}=\verb"diag" \left[ \rho_{bxx}, \rho_{bxx}
\right]$, $\rho_{f}=\verb"diag" \left[ \rho_{fxx}, \rho_{fxx}
\right],$ and in the case of drag we have in addition
$\mathbf{J}^2=0,\mathbf{J}^1-$finite. Then from
Eqs.~(\ref{eff_field_1.1})-(\ref{definicija draga}) via elementary
algebraic manipulations we obtain:
\begin{eqnarray}
\nonumber \rho^{11}=1/2\{\left[\rho_b^{-1}+\rho_f^{-1}\right]^{-1}+2\epsilon\\
+\left[(\rho_f+2\epsilon)^{-1}+\rho_b^{-1}\right]^{-1}\},\\
\nonumber
\rho^D=1/2\{\left[\rho_b^{-1}+\rho_f^{-1}\right]^{-1}+2\epsilon
\\ -\left[(\rho_f+2\epsilon)^{-1}+\rho_b^{-1}\right]^{-1}\},
\end{eqnarray}
or, in terms of matrix elements:
\begin{eqnarray}\label{prvi slucaj-matricni elementi1}
\rho_{xx}^D=-\frac{2\rho_{bxx}^2\delta^2}{(\rho_{bxx}+\rho_{fxx})^3+4(\rho_{bxx}+\rho_{fxx})\delta^2},\\
\label{prvi slucaj-matricni elementi2}
\rho_{xy}^D=\frac{\delta(2\rho_{bxx}\rho_{fxx}+\rho_{fxx}^2+4\delta^2)}{\rho_{bxx}^2+2\rho_{bxx}\rho_{fxx}+\rho_{fxx}^2+4\delta^2},\\
\label{prvi slucaj-matricni elementi3}
\nonumber \rho_{xx}^{11}=\frac{2\rho_{bxx}^{2}\delta^{2}}{(\rho_{bxx}+\rho_{fxx})^{3}+4(\rho_{bxx}+\rho_{fxx})\delta^{2}} \\
+\frac{\rho_{bxx}\rho_{fxx}}{(\rho_{bxx}+\rho_{fxx})}.
\end{eqnarray}
Formulas in Eqs.~(\ref{prvi slucaj-matricni elementi1})-(\ref{prvi
slucaj-matricni elementi3}) include parameters $\delta$,
$\rho_{bxx}$ and $\rho_{fxx}$, the last two being free parameters
about which nothing can be said \emph{a priori}. This prompted Simon
\emph{et al.}~\cite{srm03} to reason as follows. At large $d/l_B$
the number of composite bosons is small because the condensate is
broken and $\rho_{bxx}$ is large compared to $\delta$, which is the
typical Hall resistance. On the other hand, from the
experiments~\cite{willett} we know that for large $d/l_B$ holds
$\rho_{fxx}\ll \delta$. Furthermore, even as $d/l_B$ is decreased,
we expect $\rho_{fxx}$ to increase only slightly.~\cite{srm03} All
in all, for large $d/l_B$ they assume $\rho_{bxx}\gg \delta \gg
\rho_{fxx}$, and if in addition we allow $\rho_{bxx}\rho_{fxx}\ll
\delta^2$, asymptotically we obtain:
\begin{eqnarray}
\rho_{xx}^D\approx -\frac{2\delta^2}{\rho_{bxx}},\\
\rho_{xy}^D \approx 4\delta (\frac{\delta}{\rho_{bxx}})^2,\\
|\rho_{xx}^{11}| \approx |\rho_{xx}^D|.
\end{eqnarray}
Semicircle law follows directly from the previous formulas:
\begin{eqnarray}\label{semicircle}
(\rho_{xx}^D)^2+(\rho_{xy}^D-\frac{\delta}{2})^2\approx
(\frac{\delta}{2})^2,
\end{eqnarray}
in agreement with Ref.~\onlinecite{stern-halperin} (semicircle law
is of general validity for twocomponent systems in two dimensions
and it serves us as a crucial test for the the line of reasoning
quoted above, which may at first sound somewhat na\"{\i}ve).

In the opposite limit (when $d/l_B$ is reduced),  $\rho_{bxx}\ll
\rho_{fxx}\ll \delta$, because $\rho_{bxx}$ drops as a result of
Bose condensation~\cite{srm03}. When $\rho_{bxx}\rightarrow 0$ we
obtain the quantization of Coulomb drag:
\begin{eqnarray}\label{quantization of drag1}
\rho_{xy}^D\approx \delta, \\
\label{quantization of drag2} \rho_{xx}^D \longrightarrow 0,
\end{eqnarray}
as measured by Kellogg \emph{et al.}~\cite{kellogg}

Let us return now to the case of dominant intracorrelations, the
vortex metal state~\cite{milica-preprint} represented by
Eq.~(\ref{psi2}). From Fig.~\ref{fig:korelacije} the formulas for
effective fields are modified into:
\begin{eqnarray}
\mathcal{E}_{f}^{\sigma}=
\mathbf{E}^{\sigma}-2\mathbf{\epsilon}\mathbf{J}_{f}^{\sigma}
-2\mathbf{\epsilon}\mathbf{J}_b^{\sigma},\\
\mathcal{E}_{b}^{\sigma}=
\mathbf{E}^{\sigma}-\mathbf{\epsilon}(\mathbf{J}_{b}^{1}
+\mathbf{J}_b^{2}+2\mathbf{J}_f^{\sigma}),
\end{eqnarray}
and the analogous calculation yields the resistivity tensors:
\begin{eqnarray}
\nonumber \rho^{11}=\frac{1}{2}\{\left[\rho_b^{-1}+\rho_f^{-1}\right]^{-1}+2\epsilon\\
+\left[(\rho_b-2\epsilon)^{-1}+\rho_f^{-1}\right]^{-1}\left[(\rho_b-2\epsilon)^{-1}\rho_b+2\rho_f^{-1}\epsilon\right])\},
\\
\nonumber
\rho^D=\frac{1}{2}\{\left[\rho_b^{-1}+\rho_f^{-1}\right]^{-1}+2\epsilon\\
-\left[(\rho_b-2\epsilon)^{-1}+\rho_f^{-1}\right]^{-1}\left[(\rho_b-2\epsilon)^{-1}\rho_b+2\rho_f^{-1}\epsilon\right]\}.
\end{eqnarray}
The matrix elements of these tensors are:
\begin{eqnarray}
\rho_{xx}^D=-\frac{2\rho_{fxx}^2\delta^2}{(\rho_{bxx}+\rho_{fxx})^3+4(\rho_{bxx}+\rho_{fxx})\delta^2},\\
\rho_{xy}^D=\frac{\rho_{fxx}^2\delta}{(\rho_{fxx}+\rho_{bxx})^2+4\delta^2},\\
\nonumber \rho_{xx}^{11}=
\frac{2\rho_{fxx}^{2}\delta^{2}}{(\rho_{bxx}+\rho_{fxx})^3+4(\rho_{bxx}+\rho_{fxx})\delta^2}\\
+\frac{\rho_{bxx}\rho_{fxx}}{(\rho_{bxx}+\rho_{fxx})}.
\end{eqnarray}
In this case as well, there are two physically significant limits
depending on the assumptions for the values of $\rho_{bxx},
\rho_{fxx}$. In the case when $\rho_{bxx}\ll \rho_{fxx}\ll \delta$:

\begin{eqnarray}
\rho_{xx}^D \approx -\frac{\rho_{fxx}}{2},\\
\rho_{xy}^D \approx \frac{1}{4}\frac{\rho_{fxx}^2}{\delta},\\
\rho_{xx}^{11} \approx \frac{\rho_{fxx}}{2},
\end{eqnarray}
and the semicircle law follows, Eq.~(\ref{semicircle}), whereas
$|\rho_{xx}^D|=|\rho_{xx}^{11}|.$ Similarly, in the regime
$\rho_{bxx} \ll \delta \ll \rho_{fxx}$ we deduce the quantization of
Coulomb drag, Eqs.~(\ref{quantization of drag1})-(\ref{quantization of drag2}).

We emphasize that these two limits are different from those in Simon
\emph{et al.}~\cite{srm03} For example, the semicircle law was
derived assuming that $\rho_{bxx}$ is small (which is exactly the
opposite situation to the one in Ref.~\onlinecite{srm03}), while
$\rho_{fxx}$ is not necessarily small with respect to $\delta$. As
noted in the first case above, the exact values for
$\rho_{bxx},\rho_{fxx}$ are in fact unknown and this prevents us
from discriminating between the different proposed limits. In other
words, we cannot say which one of the proposed limits is plausible -
the analysis above serves us only to conclude that each of the two
composite boson-composite fermion mixed states are able (with
certain assumptions) to reproduce the phenomenology of drag
experiments.

\section{Chern-Simons theory for bilayer}\label{cs teorija za dvosloj}

Encouraged by the preliminary analysis from the previous section, we
will pursue the idea of composite boson-composite fermion mixture
further by formulating an example of Chern-Simons (CS) field theory which
can contain wavefunctions $\Psi_1,\Psi_2$ as ground states. We do
not embark on such a task only for the sake of completeness, but
also because such a theory would enable efficient calculation of
response functions and provide insight into the long range order of
the system and the nature of low-lying excitations. General drawback
of CS theories is the inability to include the projection
to LLL which is the arena where all the physics must be taking
place. Nevertheless, we will use these theories established in the works of Zhang \emph{et al.}~\cite{zhk} of composite bosons and Halperin \emph{et al.}~\cite{hlr} for composite fermions because even analyses done in advanced, projected to the LLL type of theories, in the work of Murthy and Shankar~\cite{msrev} came to the conclusion that in order to get, in the most efficient way,
to the qualitative picture of the physics of response, the usual CS theories are quite enough and accurate. In addition to this simplification, in constructing the CS theories, we will neglect the antisymmetrization requirement implied by  Eqs.~(\ref{psi1}) and (\ref{psi2}). The reason for this is that just like in hierarchical constructions, composite fermions represent
meron excitations, see Ref.~\onlinecite{milica-preprint}, that quantum disorder the $111$-state and, as it is usual when we discuss the dual picture of the FQHE~\cite{wb}, we do not extend the antisymmetrization requirement to the quasiparticle part of the electron fluid.  

Therefore we start from the lagrangian given
by~\cite{milica-preprint}:
\begin{eqnarray}
\nonumber \mathcal{L}=\sum_{\sigma}\{
\Psi_{\sigma}^{\dagger}(i\partial_0-a_0^{F\sigma}+A_0+\sigma
B_0)\Psi_{\sigma} \\
\nonumber -\frac{1}{2m}|(-i\nabla+\mathbf{a}^{F\sigma}-\mathbf{A}-\sigma\mathbf{B})\Psi_{\sigma}|^2\}\\
\nonumber + \sum_{\sigma}\{
\Phi_{\sigma}^{\dagger}(i\partial_0-a_0^{B\sigma}+A_0+\sigma
B_0)\Phi_{\sigma}\\
\nonumber -\frac{1}{2m}|(-i\nabla+\mathbf{a}^{B\sigma}-\mathbf{A}-\sigma\mathbf{B})\Phi_{\sigma}|^2\}\\
\nonumber
+\sum_{\sigma}\sum_{i=F,B}\frac{1}{2\pi}\frac{1}{2}a_0^{i\sigma}(\nabla\times
\tilde{\mathbf{a}}^{i\sigma})\\
-\frac{1}{2}\sum_{\sigma,\sigma'}\int
d^2\mathbf{r'}\delta\rho_{\sigma}(\mathbf{r})V_{\sigma\sigma'}\delta\rho_{\sigma'}(\mathbf{r'}),
\end{eqnarray}
where $\sigma$ enumerates the layers, $\Psi_{\sigma}$ i
$\Phi_{\sigma}$ are composite fermion and composite boson fields in
the layer $\sigma$,
$V_{\uparrow\uparrow}=V_{\downarrow\downarrow}\equiv V_a$,
$V_{\uparrow\downarrow}=V_{\downarrow\uparrow}\equiv V_e$, and the
densities are
$\delta\rho_{\sigma}=\delta\rho_{\sigma}^F+\delta\rho_{\sigma}^B$. By $\mathbf{A}$ (and $\mathbf{B}$) here we mean
external fields in addition to the vector potential of the uniform magnetic field, $\mathbf{A_B}$, which is accounted for and included in gauge fields
$\mathbf{a}^{F(B)\sigma}$. Therefore we have $\mathbf{a}^{F(B)\sigma}=\tilde{\mathbf{a}}^{F(B)\sigma}-\mathbf{A_B}$. External fields $\mathbf{A}$ and $\mathbf{B}$ couple with charge and
pseudospin, and in general we must introduce four gauge
fields $\mathbf{a}^{F(B)\sigma}$. Fortunately, not all of them are
independent. In the first case, the relation analogous to
Eq.~(\ref{flux-particle 1}) becomes the following gauge field
equation:
\begin{eqnarray}\label{polja 1 slucaj}
\nonumber \frac{1}{2\pi}\nabla \times
\mathbf{a}^{F\sigma}=2\delta\rho^{F\sigma}+\delta\rho^{B\uparrow}+\delta\rho^{B\downarrow},\\
\frac{1}{2\pi}\nabla \times
\mathbf{a}^{B\sigma}=\delta\rho^{F\uparrow}+\delta\rho^{F\downarrow}+\delta\rho^{B\uparrow}+\delta\rho^{B\downarrow}.
\end{eqnarray}
From the equations above, it is obvious that there are only two
linearly independent gauge fields:
$\mathbf{a}_C=\frac{\mathbf{a}^{F\uparrow}+\mathbf{a}^{F\downarrow}}{2}=\frac{\mathbf{a}^{B\uparrow}+\mathbf{a}^{B\downarrow}}{2}$
and
$\mathbf{a}_S=\frac{\mathbf{a}^{F\uparrow}-\mathbf{a}^{F\downarrow}}{2},$
and Eq.~(\ref{polja 1 slucaj}) expressed in Coulomb gauge reads:
$\frac{ika_C}{2\pi}=\delta\rho_{\uparrow}+\delta\rho_{\downarrow}\equiv\delta\rho$
and
$\frac{ika_S}{2\pi}=\delta\rho^{F\uparrow}-\delta\rho^{F\downarrow}\equiv
\delta\rho_S^{F}$ ($a_C$, $a_S$ are the transverse components of the gauge fields).
These are the constraints we wish to include into
the functional integral via Lagrange multipliers $a_0^C$ and
$a_{0}^{S}$. The interaction part of the lagrangian is easily
diagonalized by introducing $V_C=\frac{V_a+V_e}{2}$ and
$V_S=\frac{V_a-V_e}{2}.$

The strategy for integrating out the bosonic functions is the
Madelung ansatz
$\phi_{\sigma}=\sqrt{\rho_{\sigma}+\overline{\rho}_{\sigma}}e^{i\theta_{\sigma}}$,
which expands the wavefunction in terms of a product of its
amplitude and phase factor, while fermionic functions are treated as
elaborated in Ref.~\onlinecite{hlr}. After Fourier transformation,
within the quadratic (RPA) approximation, and introducing
substitutions
$\delta\rho_C^{i}=\delta\rho_{\uparrow}^{i}+\delta\rho_{\downarrow}^{i},\delta\rho_S^i=\delta\rho_{\uparrow}^i-
\delta\rho_{\downarrow}^i,i=F,B$ and
$\theta_C=\frac{\theta_{\uparrow}+\theta_{\downarrow}}{2},\theta_S=\frac{\theta_{\uparrow}-\theta_{\downarrow}}{2},$
all the terms neatly decouple into a charge and a pseudospin
channel:
\begin{eqnarray}\label{lagranzijan-charge1}
\nonumber \mathcal{L}_C=K_{00}(\delta a_0^C)^2+K_{11}(\delta
a_C)^2\\
\nonumber +i\omega\delta\rho_C^B\theta_C-\delta\rho_C^B\delta
a_0^C-\frac{\overline{\rho}_b
}{m}k^2\theta_C^2-\frac{\overline{\rho}_b}{m}(\delta a_C)^2\\
+\frac{1}{2\pi}a_0^C ik a_C-\frac{1}{2}\frac{k^2
a_C^2}{(2\pi)^2}V_C,
\end{eqnarray}
\begin{eqnarray}\label{lagranzijan-ps1}
\nonumber \mathcal{L}_{PS}=K_{00}(\delta a_0^S)^2+K_{11}(\delta
a_S)^2\\
\nonumber +i\omega\delta\rho_S^B\theta_S- \delta\rho_S B_0 -\frac{\overline{\rho}_b}{m}k^2\theta_S^2-\frac{\overline{\rho}_b}{m} B^2\\
+\frac{1}{2\pi}a_0^S ik a_S - \frac{1}{2}V_S
(\delta\rho_S^B+\frac{ik}{2\pi}a_S)^2,
\end{eqnarray}
where $\delta a_0^C\equiv a_0^C-A_0$, $\delta a_C\equiv a_C-A$,
$\delta a_0^S\equiv a_0^S-B_0$, $\delta a_S\equiv a_S-B$,
$\overline{\rho}_b$-mean density of bosons in (each) layer. In
writing down
Eqs.~(\ref{lagranzijan-charge1}),(\ref{lagranzijan-ps1}) we utilize
a compact notation suppressing $k$ ($-k$) dependence, where all the
quadratic terms of the type $\left[ X+Y \right]^2$ stand for $\left[
X(-k)+Y(-k)\right] \left[X(k)+Y(k)\right]$. $K_{00}(k)$ and
$K_{11}(k)$ are the free fermion (RPA) density-density and
current-current correlation functions.~\cite{hlr} In the long
wavelength limit ($k/k_f\ll 1$) they can be explicitly evaluated
from the general expressions~\cite{hlr}:
\begin{eqnarray}\label{fermionski odgovor_00}
\nonumber \nonumber K_{00}(k,\omega)=\frac{m}{2\pi}( 1-
\Theta(x^2-1) \frac{|x|}{\sqrt{x^2-1}}\\
+i\Theta(1-x^2)\frac{|x|}{\sqrt{1-x^2}} ),\\
\label{fermionski odgovor_11} \nonumber
K_{11}(k,\omega)=\frac{2n_f}{m}( -x^2-\frac{k^2}{24\pi
n_f}\\
\nonumber +\Theta(x^2-1)|x|\sqrt{x^2-1}\\
+i\Theta(1-x^2)|x|\sqrt{1-x^2} ),
\end{eqnarray}
where $x=\frac{m \omega}{k_f k}$, $k_f$-Fermi wavevector and
$n_f$-fermion density, $\Theta$- Heaviside step function. The mass
appearing in expressions for $K_{00}, K_{11}$ is equal to the bare
electron mass only in RPA approximation (in which we work here).

Focusing on the charge channel only,
Eq.~(\ref{lagranzijan-charge1}), and integrating out first
$\delta\rho_C^B$, then $a_{0}^{C}$ and $\delta a_C$, we arrive at
the density-density correlator:
\begin{eqnarray}\label{korelaciona funkcija gustina gustina charge kanal}
\pi_{00}(k)=\frac{(\frac{k}{2\pi})^2}{\frac{2\overline{\rho}_b}{m}-2K_{11}+V_C(\frac{k}{2\pi})^2-\frac{(\frac{k}{2\pi})^2}{\frac{2\overline{\rho}_bk^2}{m\omega^2}-2K_{00}}}.
\end{eqnarray}
In the limiting case $x\ll 1$: $K_{00} \approx \frac{m}{2\pi}(1+ix),
K_{11} \approx -\frac{k^2}{12\pi m}+i\frac{2n_f}{m}x$, and we
conclude that as $\omega\rightarrow 0$ (and then $k\rightarrow 0$) the system is
incompressible in the charge channel, so long as there is a
thermodynamically significant density of bosons $\overline{\rho}_b$.

In the pseudospin channel, we are primarily looking for the
signature of a Bose condensate i.e. whether there exists a Goldstone
mode of broken symmetry and what is the long range order of the
state. Therefore, in Eq.~(\ref{lagranzijan-ps1}) we set
$A_{\mu}=B_{\mu}=0$, and integrate over $a_0^S, a_S,
\delta\rho_S^B$:
\begin{eqnarray}\label{korelaciona funkcija teta-teta ps kanal}
\langle \theta_S(-k)\theta_S(k)\rangle =
\frac{V_S}{\omega^2\frac{\frac{1}{2}V_S+\alpha}{\alpha}-\frac{2
\overline{\rho}_b V_S}{m}k^2},
\end{eqnarray}
where $\alpha=\frac{1}{4}K_{00}^{-1}-K_{11}(\frac{2\pi}{k})^2$ (in
Appendix~\ref{linear response} we give the full linear response in
the pseudospin channel). Indeed, there exists a Goldstone mode,
albeit with a small dissipative term (which, if desired, can be
removed by pairing construction~\cite{milica-preprint}):
\begin{eqnarray}\label{goldstonova moda 1 slucaj}
\omega^0(k)=\sqrt{\frac{2\overline{\rho}_b
V_S}{m}}k-i\frac{V_S}{16 \pi^{3/2}\sqrt{n_f}}k^3.
\end{eqnarray}

Even for large $x$, it is easy to check that the pole remains at the
same value if we assume $\overline{\rho}_b\gg n_f$ (which is, in
fact, the most appropriate assumption in this case). Also, the
imaginary term disappears in this case. Such robust Goldstone mode
implies the existence of a true ODLRO and the genuine
Bose condensate. Goldstone mode $\omega^0(k)$, Eq.~(\ref{goldstonova
moda 1 slucaj}), is easily observed in Fig.~\ref{fig:goldstonova
moda 1 slucaj}, where we plotted the real part of density-density
correlation function $\pi_{00}(k)$, Eq.~(\ref{korelacione funkcije
za psi1}), in terms of parameters $Q\equiv k/k_f$ and $x \equiv
\omega/(k k_f).$ Other (fixed) parameters are: $m=l_B=1$, $d=0.5$,
$\epsilon=12.6$, $V_S=\pi d/\epsilon$,
$\overline{\rho}_b+n_f=1/(4\pi),$ $\eta=n_f/\overline{\rho}_b=1/10.$

\begin{figure}[htb]
\begin{center}
\includegraphics[width=\linewidth]{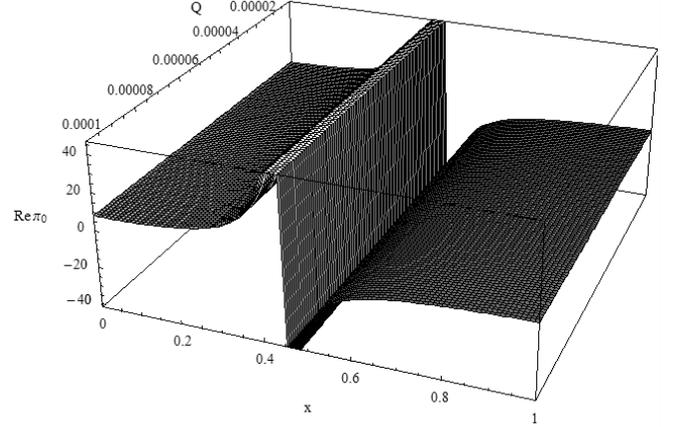}
\end{center}
\caption{$\mathrm{Re}\pi_{00}(k)$ and the Goldstone mode in the case
of $\Psi_1$} \label{fig:goldstonova moda 1 slucaj}
\end{figure}

Let us return to the second case, that of Eq.~(\ref{psi2}) and
dominant intracorrelations. According to Fig.~\ref{fig:korelacije},
relations Eq.~(\ref{polja 1 slucaj}) are modified to become:
\begin{eqnarray} \label{polja 2 slucaj}
\nonumber \frac{1}{2\pi}\nabla \times
\mathbf{a}^{F\sigma}=2\delta\rho^{F\sigma}+2\delta\rho^{B\sigma},\\
\frac{1}{2\pi}\nabla \times
\mathbf{a}^{B\sigma}=2\delta\rho^{F\sigma}+\delta\rho^{B\uparrow}+\delta\rho^{B\downarrow}.
\end{eqnarray}
It is obvious that in this case we have only $3$ linearly
independent gauge fields, namely
$a_C=\frac{a^{F\uparrow}+a^{F\downarrow}}{2}=\frac{a^{B\uparrow}+a^{B\downarrow}}{2}$,
$a_S=\frac{a^{F\uparrow}-a^{F\downarrow}}{2}$ and
$a_{FS}=\frac{a^{B\uparrow}-a^{B\downarrow}}{2}.$ Introducing the
same substitutions as before, the lagrangian again decouples into a
charge channel:
\begin{eqnarray}\label{lagranzijan charge channel psi2}
\nonumber \mathcal{L}_C=K_{00}(\delta a_0^C)^2+K_{11}(\delta
a_C)^2\\
\nonumber +i\omega\delta\rho_C^B \theta_C - \delta\rho_C^B \delta
a_0^C-\frac{\overline{\rho}_b}{m}k^2\theta_C^2-\frac{\overline{\rho}_b}{m}(\delta
a_C)^2 \\
+\frac{ik}{2\pi}a_0^C a_C - \frac{1}{2}V_C (\frac{k}{2\pi})^2 a_C^2,
\end{eqnarray}
and a pseudospin channel:
\begin{eqnarray}\label{lagranzijan ps channel psi2}
\nonumber \mathcal{L}_{PS}=K_{00}(\delta a_0^S)^2+K_{11}(\delta
a_S)^2\\
\nonumber +i\omega\delta\rho_S^B\theta_S-\delta\rho_S^B\delta
a_0^{FS}-\frac{\overline{\rho}_b}{m}k^2\theta_S^2-\frac{\overline{\rho}_b}{m}(\delta
a_{FS})^2\\
+\frac{ik}{2\pi}a_0^S a_{FS}+\frac{ik}{2\pi}a_0^{FS}(a_S-a_{FS})
-\frac{1}{2}V_S (\frac{k}{2\pi})^2a_S^2,
\end{eqnarray}
where $\delta a_0^{FS}\equiv a_0^{FS}-B_0,\delta a_{FS}=a_{FS}-B$,
all the other symbols have retained their meanings.

This time we will not analyze the charge channel in detail. To this
end, we note that the system in incompressible in this sector, the
fact which is easily established by integrating out all the gauge
fields, densities and boson phase in Eq.~(\ref{lagranzijan charge
channel psi2}).

\begin{figure}[htb]
\begin{center}
\includegraphics[width=\linewidth]{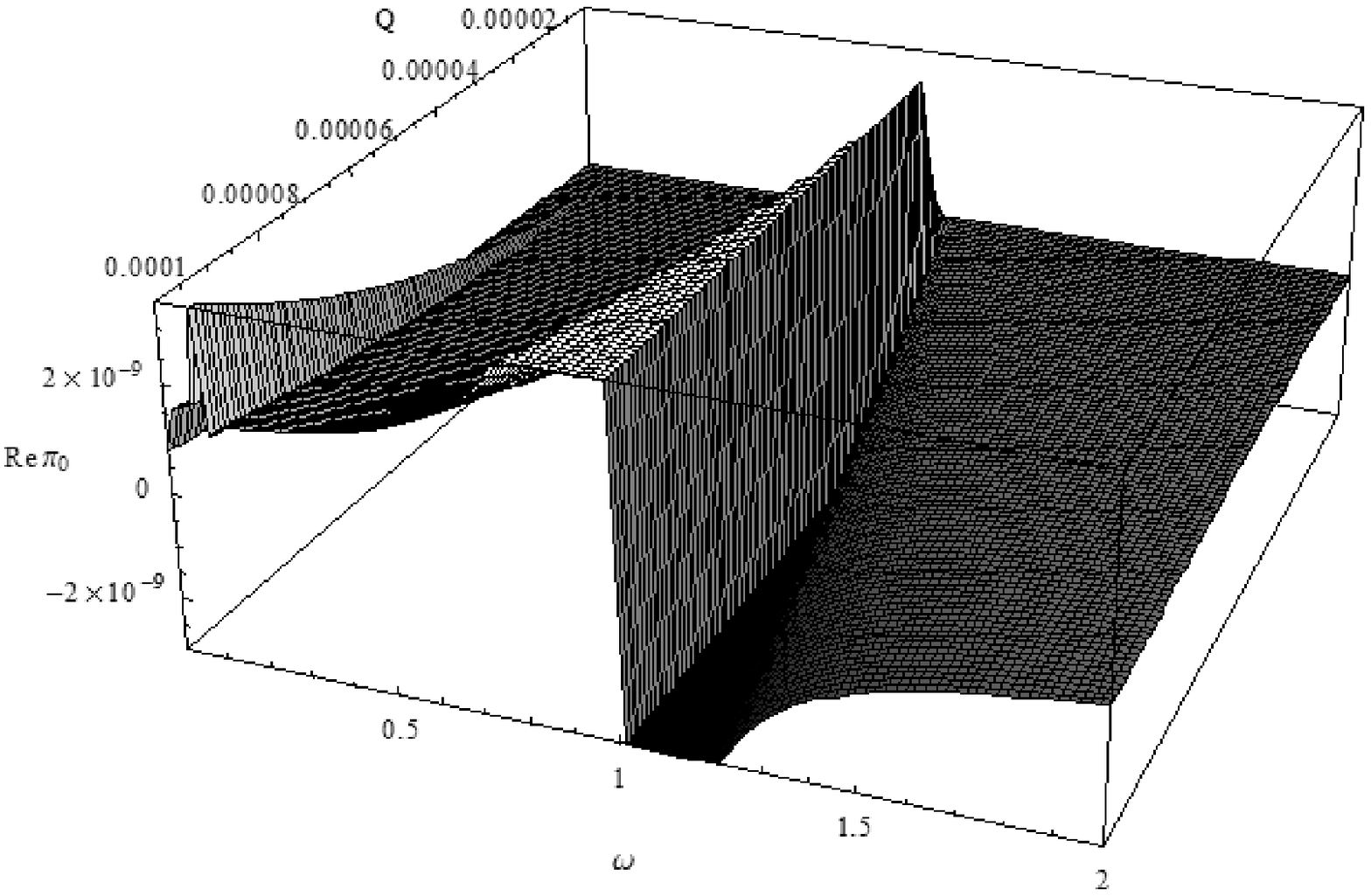}
\end{center}
\caption{$\mathrm{Re}\pi_{00}(k)$ in the case of $\Psi_2$}
\label{fig:goldstonova moda 2 slucaj}
\end{figure}

In the pseudospin channel, a calculation of the density-density correlator leads to the conclusion that in this channel the system is compressible (see also Fig.~\ref{fig:goldstonova moda 2 slucaj}). The $\theta-\theta$ correlator is:
\begin{eqnarray}
\langle \theta_S(-k)\theta_S(k)
\rangle=\frac{\frac{1}{k^2}\beta\gamma}{(\frac{\omega}{2\pi})^2(\beta+\gamma)-\frac{\overline{2\rho}_b}{m}\beta\gamma},
\end{eqnarray}
where $\beta=\frac{1}{2K_{00}}(\frac{k}{2\pi})^2+\frac{2
\overline{\rho}_b}{m}$, $\gamma=V_S(\frac{k}{2\pi})^2-2K_{11}.$ For
small $k/k_f$ and $x$, the correlator diverges for
$\omega^0=\frac{4\pi\overline{\rho}_b}{m}=const$, which obviously
contradicts the original assumption for the range of $x$ and hence
we reject this pole. For $x\gg 1$ (and still $k\ll k_f$), the
relations Eq.~(\ref{fermionski odgovor_00}),(\ref{fermionski
odgovor_11}) are approximately $K_{00} \approx -\frac{1}{4\pi x^2},
K_{11}\approx -\frac{n_f}{m}$, and we obtain two poles:
\begin{eqnarray}\label{plasma polovi1}
\omega^0(k)=\frac{4\pi n_f}{m}\sqrt{\frac{1}{2}+\eta-\frac{1}{2}\sqrt{1+4\eta}},\\
\label{plasma polovi2} \Omega^0(k)=\frac{4\pi
n_f}{m}\sqrt{\frac{1}{2}+\eta+\frac{1}{2}\sqrt{1+4\eta}},
\end{eqnarray}
where $\eta=\overline{\rho}_b/n_f$ is the ratio of boson to fermion
density (Eqs.~(\ref{plasma polovi1}),(\ref{plasma polovi2}) hold for
any $\eta$, although in the physical limit that we are presently
interested in, $\eta$ may be regarded as small). In
Fig.~\ref{fig:goldstonova moda 2 slucaj} we plotted the real part of
the density-density correlation function in the case of $\Psi_2$,
Eq.~(\ref{korelacione funkcije za psi2}). In contrast to
Fig.~\ref{fig:goldstonova moda 1 slucaj}, here we opt for $\omega$
and $Q$ as free parameters and set $d=1.5$ and
$\eta=\overline{\rho}_b/n_f=1/10$ as the more likely values in this
case. Distinctive feature of Fig.~\ref{fig:goldstonova moda 2
slucaj} at $\omega\cong 1$ is the plasma frequency $\Omega^0$ and
the smaller singularity at $\omega\cong 1/10$ corresponds to
$\omega^0$. There is also a striking absence of Goldstone mode in
this case.

We now proceed to calculate ODLRO in the pseudospin channel of
$\Psi_2$. As it turns out, ODLRO will be nontrivially modified and
assume algebraic form. We know that interaction does not affect the
value of characteristic exponent~\cite{zhang} and therefore set
$V_S\equiv 0$. Bearing in mind that we work in the long wavelength
limit, we arrive at the following expression for the correlator:
\begin{eqnarray}
\langle \theta_S(-k)\theta_S(k)
\rangle=\frac{(2\pi\omega_{P}/k^2)\left[\omega^2-\omega_P^2\eta\right]}{\left[\omega^2-
(\omega^0(k))^2\right] \left[ \omega^2-(\Omega^0(k))^2 \right]},
\end{eqnarray}
where we introduced $\omega_P=\frac{4\pi n_f}{m}$. After contour
integration over $\omega$~\cite{zhang}:
\begin{eqnarray}
\nonumber \langle \theta_S(-k)\theta_S(k) \rangle=
-\frac{2\pi}{k^2}f(\eta)
\end{eqnarray}
where $f(\eta)=\frac{1}{\sqrt{1+4\eta}},$ which leads to the
algebraic ODLRO:
\begin{eqnarray}\label{long range order psi2}
\nonumber \langle
e^{i\theta_S(\mathbf{r})}e^{-i\theta_S(\mathbf{r'})}\rangle \propto
\frac{1}{|\mathbf{r}-\mathbf{r'}|^{f(\eta)}}
\\ \approx |\mathbf{r}-\mathbf{r'}|^{-(1-2\eta+o(\eta^2))}.
\end{eqnarray}

This algebraic ODLRO persists as long as $\eta>0$ (function $f$ is
positive everywhere in this domain). The expression Eq.~(\ref{long
range order psi2}) is formally reminiscent of BKT XY-ordering, only
the role of the temperature is overtaken by the parameter $\eta$
(the analysis of this paper assumes temperature $T=0$). Pursuing
this analogy further, we conclude that the relative fluctuations of
composite boson and composite fermion density represent the
mechanism which may lead to the ultimate breakdown of the
$111$-condensate.

\section{Evolution of the ground state with $d$}\label{evolution}

\begin{figure}[htb]
\begin{center}
\includegraphics[width=\linewidth]{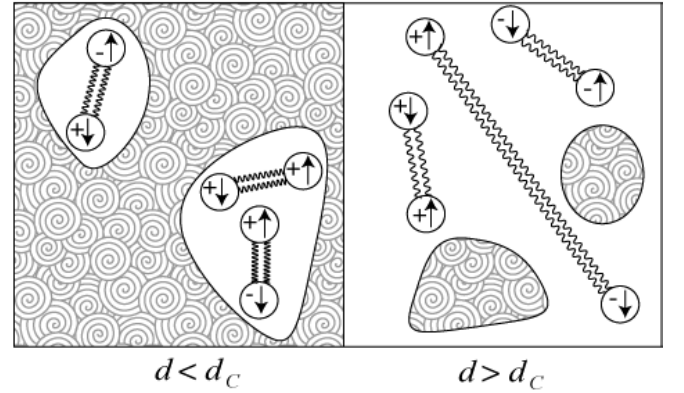}
\end{center}
\caption{Evolution of the ground state with varying $d$, before and
after the transition at $d = d_{C}$. The regions with meron pairs
represent the vortex metal ($\Psi_{2}$) phase. The background
represents the superfluid ($\Psi_{1}$) phase.} \label{fig:evolution}
\end{figure}

In order to investigate the transition from the incompressible,
$111$-like state at lower $d/l_{B}$, to the compressible, possibly
vortex metal-like state at higher $d/l_{B}$, we are motivated to
introduce what we call \emph{generalized vortex metal}. In addition
to the ordinary vortex metal ($\Psi_{2}$), we include (in each
layer) another kind of composite fermions that connect to the
composite boson sea as in the case of $\Psi_{1}$. The generalized vortex metal is clearly the only additional option
left of connecting electrons divided in composite bosons and composite fermions beside the two extreme cases, $\Psi_{1}$ and $\Psi_{2}$. Again, in this state some of composite fermions connect in the manner of $111$-state to the composite bosons and the rest of composite fermions connect exclusively to the composite bosons of the same layer in the manner of the Rezayi-Read state. This is succinctly represented by the following gauge field constraints:
\begin{eqnarray}
\frac{1}{2\pi} \nabla \times
a^{B\sigma}=\delta\rho_{B\uparrow}+\delta\rho_{B\downarrow}+\delta\rho_{F\uparrow}^{(1)}+\delta\rho_{F\downarrow}^{(1)}+2\delta\rho_{F\sigma}^{(2)},\\
\frac{1}{2\pi}\nabla\times a_{1}^{F\sigma}=\delta\rho_{B\uparrow}+\delta\rho_{B\downarrow}+2\delta\rho_{F\sigma}^{(1)}+2\delta\rho_{F\sigma}^{(2)},\\
\frac{1}{2\pi}\nabla\times
a_{2}^{F\sigma}=2\delta\rho_{B\sigma}+2\delta\rho_{F\sigma}^{(1)}+2\delta\rho_{F\sigma}^{(2)},
\end{eqnarray}
where the superscripts $(1),(2)$ indicate composite fermion species
in each layer. Chern-Simons theory easily follows from the above
gauge field equations and yields incompressible behavior in the
charge channel. In the pseudospin channel:
\begin{eqnarray}\label{lagrangian gen vortex metal}
\nonumber \mathcal{L}_{PS}=K_{00}^{(1)}(\delta a_{0,1}^{FS})^2+K_{00}^{(2)}(\delta a_{0,2}^{FS})^2\\
\nonumber +K_{11}^{(1)}(\delta a_{1}^{FS})^2+K_{11}^{(2)}(\delta a_{2}^{FS})^2 \\
\nonumber +i\omega\delta\rho_{S}^{B}\theta_{S}-\delta\rho_{S}^{B}
\delta a_{0}^{S}-\frac{\overline{\rho}_b}{m}
k^2\theta_{S}^{2}-\frac{\overline{\rho}_b}{m}(\delta a_{S})^{2}
\\
\nonumber
+\frac{ik}{2\pi}a_{0,1}^{FS}(a_{1}^{FS}-a_{S})+\frac{ik}{2\pi}a_{0}^{S}(a_{2}^{FS}-a_{1}^{FS})\\
\nonumber +\frac{ik}{2\pi}a_{0,2}^{FS}a_{S}-\frac{1}{2}V_{S}(\frac{k}{2\pi})^{2}|a_{2}^{FS}|^{2}\\
-\frac{1}{2}V_{hc}(\frac{k}{2\pi})^{2}a_{S}(a_{1}^{FS}-a_{S}),
\end{eqnarray}
where the linearly independent fields are given by
$a_{S}=\frac{a^{B\uparrow}-a^{B\downarrow}}{2}$,
$a_{1}^{FS}=\frac{a_{1}^{F\uparrow}-a_{1}^{F\downarrow}}{2}$,
$a_{2}^{FS}=\frac{a_{2}^{F\uparrow}-a_{2}^{F\downarrow}}{2}$,
subscripts $1,2$ distinguish between composite fermion species and
$^{S}$ denotes antisymmetric combination of the densities in two
layers (like in Sec.~\ref{cs teorija za dvosloj}). A noteworthy
feature of the lagrangian, Eq.~(\ref{lagrangian gen vortex metal}),
is the existence of $V_{hc}$, the hard-core repulsion term between
the two species of composite fermions inside each layer. The
presence of such a term (added by hand) is natural if we imagine
composite fermions residing in two separate Fermi spheres. However,
the danger of blindly introducing this term is that it may
incidentally bring about the incompressible behavior (otherwise not
present) in the system. We have verified that this is \emph{not} the
case here i.e. the system remains incompressible whether or not we
choose to introduce $V_{hc}$. It therefore appears more intuitive to
keep $V_{hc}$, taking the limit $V_{hc}\rightarrow \infty$ in the
end. Step by step, eliminating all the gauge fields, we are lead to
the following correlation function:
\begin{eqnarray}
\langle \theta_{S}(-k)\theta_{S}(k) \rangle =
\frac{V_{S}+\frac{2n_{f2}}{m}(\frac{2\pi}{k})^{2}}{\omega^{2}-(\frac{2\overline{\rho}_{b}V_{S}}{m}k^2+(\frac{4\pi}{m})^{2}\overline{\rho}_{b}n_{f2})}
\end{eqnarray}
and the low-energy spectrum is dominated by the plasma frequency:
\begin{eqnarray}
\omega^{0}(k)=\frac{4\pi}{m}\sqrt{\overline{\rho}_{b}n_{f2}},
\end{eqnarray}
where $n_{f2}$ is the density of the composite fermions which bind
exclusively within the layer they belong to. Generalized vortex
metal therefore is a state that only supports gapped collective
excitations, despite the presence of composite bosons and the kind
of composite fermions which enforce interlayer correlation. If it is
pertinent to the region  of the tunneling experiments of Spielman
\emph{et al.}~\cite{spielman} and counterflow experiments of Kellogg
\emph{et al.}~\cite{kellogg}, we believe that our homogeneous theory
of Sec.~\ref{cs teorija za dvosloj},\ref{evolution} then suggests
that (generalized) vortex metal can appear only as localized islands
(due to presence of disorder at low temperatures) amidst the
background of $\Psi_{1}$ phase (Fig.~\ref{fig:evolution}). In
Fig.~\ref{fig:evolution} depicted are weakly-coupled
vortex-antivortex pairs i.e. meron-antimeron pairs (due to the
charge degree of freedom, there are four kinds of merons
\cite{kmoon}) inside the vortex metal phase. They are expected to
exist in the vortex-metal phase on the grounds of disordering of the
correlated phase. As argued in Ref.~\onlinecite{milica-preprint},
the inclusion of composite fermions into the $111$ state ($\Psi_{1}$
and $\Psi_{2}$) corresponds to creation of meron-antimeron pairs.
There are more pairs and more of larger size as $d$ increases
consistent with the BKT picture of the phase that supports algebraic
ODLRO, Eq.~(\ref{long range order psi2}).

\section{Further comparison with experiments} \label{further comparison}

In this section we wish to address in depth the potential of the
model states, $\Psi_{1}$ and $\Psi_{2}$, in explaining the
phenomenology of experiments on bilayer. The key question in this
analysis is: what is the nature of the compressible phase
corresponding to higher $d/l_{B}$ that still harbors some of the
intercorrelation present at lower $d/l_{B}$~\cite{kelloggdrag}?

The answer to this question cannot be given by looking at simple
transport properties. In Sec.~\ref{sect:basic chern-simons} it was
shown that both $\Psi_{1}$ and $\Psi_{2}$ in certain regimes can
recover the two main experimental findings of Kellogg \emph{et al.}
in drag experiments: the semicircle law~\cite{kelloggdrag} and the
quantization of Hall drag resistance~\cite{kellogg3}. On the other
hand, our Chern-Simons RPA approach at $T=0$ stresses that all that
states considered in this paper are incompressible. However, at
finite $T$, a finite energy~\cite{milica-preprint} is needed to
excite a meron in $\Psi_{2}$ and therefore $\Psi_{2}$ seems like a
better candidate for exhibiting compressible behavior at any finite
$T$ or, at least, a very small gap. Furthermore, within the vortex
metal picture, $\Psi_{2}$ allows the following simple scenario. For
$\rho_{bxx} \ll \rho_{fxx} \ll \delta$, one gets the semicircle law
derived in Sec.~\ref{sect:basic chern-simons}. As $d/l_{B}$
increases, the density of bosons decreases and one enters the regime
$\rho_{fxx}\ll \rho_{bxx} \ll \delta$ where $|\rho_{xx}^{D}|\gg
\rho_{xy}^{D}$ (as witnessed in the experiments~\cite{kelloggdrag}).
The persistence of enhanced longitudinal drag
resistance~\cite{kelloggdrag} up to very high $d/l_{B}$ provides
additional support to our choice of $\Psi_{2}$ which can explain the
remaining intercorrelation (drag) in the case where explicit
tunneling is absent. Finally, as $\rho_{bxx}\rightarrow \infty$,
both resistances go to zero, the bilayer decouples and bosons vanish
from the system.

Our picture is certainly incomplete because it does not explicitly
include the effects of disorder (which must be very relevant for the
physics of bilayer in the regimes $d\sim l_B$ - a simple way to see
this is to look at the behavior of measured
counterflow resistances~\cite{kellogg,tutuc} $\rho_{xx}^{CF}$, $\rho_{xy}^{CF}$
that enter the insulating regime very quickly after passing through
$\nu_{T}=1$). Fertig and Murthy~\cite{fertig} provided a realistic
model for the effects of disorder and in their disorder-induced
coherence network in the incompressible phase of the bilayer, merons
are able to sweep by hopping  across the system, causing the
activated behavior of resistance (dissipation) in counterflow. This
finding is consistent with our own.

At the end our picture is in the spirit of the Stern and Halperin
proposal~\cite{stern-halperin} but instead of the $1/2$ compressible
phase coexisting in a phase separated picture with the superfluid
phase ($\Psi_{1}$), we assume the existence of the vortex metal
phase ($\Psi_{2}$). This coincides with the Fertig and Murthy
proposal~\cite{fertig} for the incompressible region that explains
the "imperfect" superfluid behavior. It is the continuous
extrapolation of this phase separated picture that brings and favors
$\Psi_{2}$ for larger $d/l_{B}$ (instead of $\Psi_{1}$). There
$\Psi_{2}$ is able to explain the persistence of intercorrelations
through enhanced longitudinal drag accompanied by the absence of
tunneling and phase coherence.~\cite{kellogg3}

Finally, we are able to account for the effects of the layer density
imbalance in tunneling, drag~\cite{spielman-imbalance} and
counterflow~\cite{tutuc-imbalance} experiments. Spielman \emph{et
al.}~\cite{spielman-imbalance} observed that small density imbalance
stabilizes the resonant tunneling peak - a simple reason for this is
that $\Psi_{1}$ can easily accommodate the fluctuations in density
(see comment after Eq.(\ref{flux-particle 1})). Because of the same
reason, Hall drag resistance remains quantized up to larger
$d/l_{B}$ in the presence of density imbalance. On the other hand,
the enhancement of longitudinal drag resistance at large $d/l_{B}$
was also reported~\cite{kelloggdrag} to be insensitive to density
imbalance. While the reason for this cannot be seen only from
looking at the form of $\Psi_{2}$ (this state constrains both
fermion and boson numbers in two layers, see comment after
Eq.~\ref{flux-particle 2}), we believe that meron excitations are
responsible for absorbing the density fluctuations, especially at
finite $T$.

Recently, the quantum Hall bilayer was probed using resonant
Rayleigh scattering~\cite{rrs} for samples with different tunneling
amplitudes and when the in-plane magnetic field is present. They
detected a nonuniform spatial structure in the vicinity of the
transition, suggesting a phase-separated version of the ground
state. Our results (for zero tunneling limit and excluding disorder)
hint that such phase separation may indeed be necessary to invoke in
order to achieve a full description of the strongly-coupled,
incompressible phase and the transition in a bilayer.

\section{Discussion and Conclusion}\label{conclusion}

In conclusion we showed how two model states, $\Psi_{1}$ and
$\Psi_{2}$, can account for the basic phenomenology of the bilayer
that came up from various experiments.

A very interesting question pertains the model state $\Psi_{2}$. As
effectively a state that represents a collection of meron
excitations interacting through topological interactions, a question
comes when they are in a confined (dipole) phase and when in a
metallic (plasma) phase. So in principle we can expect that the
static correlator in Eq.~(\ref{long range order psi2}) can be
reproduced by considering a 2D bosonic model with meron excitations
interacting via 2D Coulomb plasma interaction.~\cite{tsvelik}
Therefore we believe that the Laughlin ansatz~\cite{laughlin} of
considering (static) ground state correlators as statistical models
in 2D can be applied here also. We expect that the ground state
correlators in a dual approach, in which we switch from composite
fermion to meron coordinates, can be mapped to a partition function
of a 2D Coulomb plasma.~\cite{polyakov} The 2D Coulomb plasma has
two different phases. For large $\beta$ (inverse $T$) the charges
form dipoles and the system is with long range correlations (no mass
gap). At some critical $\beta$, dissociation of dipoles occurs and
we have a plasma phase with a Debye screening, and therefore a mass
gap. Thus calculations that will capture more of the meron
contribution than our RPA approach in the Chern-Simons theory may
find a transition and exponential decay of the correlator
(Eq.~(\ref{long range order psi2})), before reaching
$\overline{\rho}_{b} = 0$ limit. Indeed our ODLRO exponent in
Eq.~(\ref{long range order psi2}) at $\overline{\rho}_{b} = 0$ is
$1$ which is well above the exponent of the BKT transition or
critical exponent $1/4$. At that point our system may develop a gap
in the pseudospin channel and completely lose interlayer coherence
(exponential decay of correlators). Furthermore, we expect that the
superfluid portion of the composite boson density will disappear
leading to compressible behavior in the charge channel.~\cite{zhang}
This is all consistent with experiments~\cite{kellogg3,kelloggdrag}
which find that, at $d/l_{B} \approx 1.84$, the vanishing of the
conventional quantum Hall effect and the system's Josephson-like
tunneling characteristics occurs simultaneously. Intercorrelated
bosons continue to exist without a superfluid property and lead to
enhanced $\nu = 1$ drag at large $d/l_{B}$. They disappear from the
system around $d/l_{B} \approx 2.6$~\cite{kelloggdrag}.

\section{Acknowledgment}
The work was supported by Grant No. 141035 of the  Ministry of
Science of the Republic of Serbia.

\appendix
\section{}\label{linear response}

In order to extract the response functions in functional integral
formalism, one needs to integrate over all degrees of freedom
except those of the external fields. The integration of these
fields in the RPA approximation proverbially reduces to the
Gaussian integral:
\begin{eqnarray}\label{gausov integral}
\nonumber \int d(z,z*) \verb"exp" \left[ -z* w z +u* z + v
z*\right]\\ \nonumber =\frac{\pi}{w}\verb"exp"\left[ \frac{u*
v}{w}\right], \mathrm{Re} w>0.
\end{eqnarray}
For the pseudospin channel in the case of $\Psi_1$
(Eq.~(\ref{psi1})) we therefore obtain the following linear
response:
\begin{eqnarray}\label{korelacione funkcije za psi1}
\pi_{00}(k)=\frac{1}{\Xi}(\Omega(3 V_S+2\alpha)+1),\\
\pi_{01}(k)=\pi_{10}(k)=\frac{1}{\Xi}\frac{-i \pi}{k}K_{11}(1+\Omega
V_S),\\
\nonumber \pi_{11}(k)=\frac{1}{\Xi}\{ 2V_{S}(1-\Omega
V_{S})(K_{11}-\frac{\overline{\rho}_b}{m})\\
+K_{11}/K_{00}-4\alpha
\frac{\overline{\rho}_b}{m} \},
\end{eqnarray}
where $\alpha \equiv
\frac{1}{4}K_{00}^{-1}-K_{11}(\frac{2\pi}{k})^2,$ $\Omega\equiv
\left[ V_S - \frac{m\omega^2}{2 \overline{\rho}_b k^2}\right]^{-1}$
and $\Xi=V_S(1-\Omega V_S)+2\alpha$.

The response functions in the case of the pseudospin channel of
$\Psi_2$ (Eq.~(\ref{psi2})) are:
\begin{eqnarray} \label{korelacione funkcije za psi2}
\pi_{00}=\frac{1}{\Delta}(\frac{k}{2\pi})^2,\\
\pi_{01}=\pi_{10}=\frac{1}{\Delta}\frac{ik}{2\pi}\Lambda,\\
\pi_{11}=\frac{1}{\Delta}\{ \Lambda^2+\Delta
(2K_{11}-\frac{2\overline{\rho}_b}{m}-16W^4 (\frac{2\pi
\overline{\rho}_b}{m\omega})^4) \},
\end{eqnarray}
where $W^4 \equiv -\frac{(\frac{m\omega^2}{2\overline{\rho}_b
(2\pi)^2})^2}{\frac{1}{2K_{00}}(\frac{k}{2\pi})^2-\frac{m\omega^2}{2\overline{\rho}_b
(2\pi)^2}+\frac{2\overline{\rho}_b}{m}},$ $\Delta \equiv
W^4-\frac{m\omega^2}{2\overline{\rho}_b (2\pi)^2}-2K_{11}+V_S
(\frac{k}{2\pi})^2,$ $\Lambda \equiv 4W^4(\frac{2\pi
\overline{\rho}_b}{m\omega})^2-2K_{11}.$


\begin{references}
\bibitem{halperin} B.I. Halperin, Helv. Phys. Acta {\bf 56}, 75
(1983).
\bibitem{kmoon} K. Moon, H. Mori, K. Yang, S.M. Girvin, A.H. MacDonald, L. Zheng,
D. Yoshioka, and S.-C. Zhang, Phys. Rev. B {\bf 51}, 5138 (1995).
\bibitem{macdonald} H.A. Fertig, Phys. Rev. B {\bf 40}, 1087 (1989).
%A.H.MacDonald, Physica B (Amsterdam) {\bf 298}, 129 (2001).
\bibitem{stanic} I. Stani\'{c} and M.V. Milovanovi\'{c}, Phys. Rev. B {\bf 71}, 035329 (2005).
\bibitem{spielman} I.B. Spielman, J.P. Eisenstein, L.N. Pfeiffer, and K.W. West,
Phys. Rev. Lett. {\bf 84}, 5808 (2000); {\em ibid} {\bf 87},
036803 (2001).
\bibitem{kellogg} M. Kellogg, J.P. Eisenstein, L.N. Pfeiffer, and K.W. West,
Phys. Rev. Lett. {\bf 93}, 036801 (2004).
\bibitem{tutuc} E. Tutuc, M. Shayegan, and D.A. Huse, Phys. Rev. Lett. {\bf 93}, 036802 (2004).
\bibitem{srm03} S.H. Simon, E.H. Rezayi, and M.V. Milovanovi\'{c}, Phys. Rev. Lett.
{\bf 91}, 046803 (2003).
\bibitem{kelloggdrag} M. Kellogg, J.P. Eisenstein, L.N. Pfeiffer, and K.W. West,
Phys. Rev. Lett. {\bf 90}, 246801 (2003).
\bibitem{milica-preprint} M.V. Milovanovi\'{c}, Phys. Rev. B {\bf 75}, 035314 (2007).
\bibitem{laughlin-orig} R.B. Laughlin, Phys. Rev. Lett. {\bf 50}, 1395 (1983).
\bibitem{rezayi-read} E.H. Rezayi and N. Read, Phys. Rev. Lett. {\bf 72}, 900 (1994).
\bibitem{stern-halperin} A. Stern and B.I. Halperin, Phys. Rev. Lett. {\bf 88}, 106801 (2002).
\bibitem{willett} R.L. Willett, Adv. Phys. {\bf 46}, 447 (1997).
\bibitem{zhk} S.C. Zhang,  T. H. Hansson, and S. Kivelson, Phys. Rev. Lett. {\bf 62}, 980 (1989).
\bibitem{hlr} B.I. Halperin, P.A. Lee, and N. Read, Phys. Rev. B {\bf 47}, 7312 (1993).
\bibitem{msrev} G. Murthy and R. Shankar, Rev. Mod. Phys. {\bf 75}, 1101 (2003).
\bibitem{wb} B. Blok and X.-G. Wen, Phys. Rev. B {\bf 42}, 8145 (1990); \emph{ibid.} {\bf 43}, 8337 (1991).
\bibitem{zhang} S.-C. Zhang, Int. J. of Mod. Phys. B {\bf 6}, 25 (1992).
\bibitem{kellogg3} M. Kellogg, J.B. Spielman, J.P. Eisenstein, L.N. Pfeiffer, and K.W. West,
Phys. Rev. Lett. {\bf 88}, 126804 (2002).
\bibitem{fertig} H.A. Fertig and G. Murthy, Phys. Rev. Lett. {\bf
95}, 156802 (2005).
\bibitem{spielman-imbalance} I.B. Spielman, M. Kellogg,
J.P. Eisenstein, L.N. Pfeiffer and K.W. West, Phys. Rev. B {\bf 70},
081303(R), 2004.
\bibitem{tutuc-imbalance} E. Tutuc and M. Shayegan, Phys. Rev. B {\bf
72}, 081307(R), (2005).
\bibitem{rrs} S. Luin, V. Pellegrini, A. Pinczuk, B.S. Dennis,
L.N. Pfeiffer and K.W. West, Phys. Rev. Lett. {\bf 97}, 216802 (2006).
\bibitem{tsvelik} A.O. Gogolin, A.A. Nersesyan, and A.M. Tsvelik,
{\em Bosonization and Strongly Correlated Systems} (Cambridge
University Press, Cambridge, 1998)
\bibitem{laughlin} R.B. Laughlin in {\em The Quantum Hall Effect}, 2nd. ed., edited by
R.E. Prange and S.M. Girvin (Springer, New York, 1990).
\bibitem{polyakov} A.M. Polyakov, {\em Gauge Fields and
Strings} (Harwood Academic Publishers, Chur, 1989)
\end{references}
\end{document}